**Title: The Virtual Hearing Clinic (VHC)- a modular online platform for hearing research and hearing health care**

**Authors:** Lena Schell-Majoor[a,b,f*], Kim M. Rullmann[c,d,f], Tobias Bruns[e,f], Theresa Jansen[d,f], Volker Hohmann[c,d,f], Hendrik Kayser[c,d,f], Birger Kollmeier[a,d,e,f]

[a]*Medical Physics, Universität Oldenburg, 26111 Oldenburg, Germany*
[b]*Big Data in Medicine, Universität Oldenburg, 26111 Oldenburg, Germany*
[c]*Auditory Signal Processing and Hearing Devices, Universität Oldenburg, 26111 Oldenburg, Germany*
[d] *Hörzentrum Oldenburg gGmbH, D-26129 Oldenburg, Germany*
[e] *Fraunhofer Institute for Media Technology, 26129 Oldenburg, Germany*
[f]*Cluster of Excellence Hearing4all, Universität Oldenburg, D-26111 Oldenburg, Germany*

Contact: Lena Schell-Majoor (lena.schell-majoor@uni-oldenburg.de)

## Abstract

**Objective:** The aim is to introduce the concept of the Virtual Hearing Clinic (VHC), give an overview of the current status and exemplify its feasibility with data from a diagnostics module obtaining hearing thresholds.

**Design:** The architecture of the VHC is described and an overview of functional modules that have been developed and tested in respective studies is given. As a functional example data from an experiment is presented. Hearing thresholds were obtained from 20 subjects with hearing loss with the VHC using the Graded Response Bracketing (GraBr) procedure and compared to reference thresholds obtained with a clinical audiometer.

**Results:** Median VHC-based hearing thresholds over all frequencies did not differ significantly from the reference with values of 57.4 dB SPL (VHC) and 55.5 dB SPL (reference). The shape of the frequency-dependent thresholds was also found to be very similar for 250 Hz to 4 kHz. Results for 6 kHz showed larger differences.

**Conclusion:** The VHC is suitable as a mobile hearing health application and to collect data for audiological research. By offering flexibility in time and location it can lower barriers for hearing health care and enable collecting large datasets that are needed for the advancement of data-driven audiology.

## 1. Introduction

Hearing loss is a health condition affecting a large number of people and its prevalence as well as its severity increases with age (Von Gablenz & Holube, 2016; Wilson et al., 2017; *World Report on Hearing*, 2021). This often soft onset of hearing loss can lead to some degree of habituation and, consequently, to late diagnosis and interventions, which could even be boosted by the stigma connected to hearing loss. Therefore, a relevant proportion of people that would likely benefit from using hearing aids do not own or use any (Döge et al., 2023). However, early diagnosis and intervention is important for the success of rehabilitative measures, e.g., the provision of hearing aids, and can contribute to reducing the impact and burden of hearing loss on people's lives, health and society (Wilson et al., 2017; *World Report on Hearing*, 2021). Despite significant advances in the treatment of hearing loss over the last decades, the benefit of hearing loss treatment still is highly variable and depends on several factors. This illustrates the need for early and more targeted hearing health care (Hohmann, 2023).

One approach to lower the barrier to early hearing health care is to use self-administered and web-based applications, as they are a rather low-cost, low-effort approach to improve the accessibility to basic diagnostics for people hesitant to use medical services or without access to local medical infrastructure. They can also empower and enable patients to steer their hearing health journey with a more profound knowledge of their own hearing deficits and preferences. Additionally, they provide the potential to reduce costs by, e.g., reducing the need of in-person visits and supervision by clinical staff.

Self-administered web-based approaches are subject of current audiologic research and also already used in audiologic practice, e.g., to improve the accessibility of diagnostics and outcome of hearing aid fitting. Most of the available applications focus on only one aspect, i.e., diagnostics, basic hearing support, hearing aid operation and adjustment or outcome assessments (Laird et al., 2024). The majority of diagnostic applications target hearing screening, i.e., a binary classification to assess if individual hearing is impaired or not (Frisby et al., 2022; Handzel & Franck, 2021; Oremule et al., 2024; Wang et al., 2023), or basic diagnostics usually estimating an audiogram (Al-Maskari et al., 2026; Mui et al., 2025; Swords et al., 2024; Vercammen & Strelcyk, 2025). Other available applications provide interfaces for hearing aid adjustment, basic hearing support using a mobile device with headphones or assess hearing aid outcomes also using Ecological Momentary Assessments (EMA) (Fourie et al., 2024; Jo et al., 2023; Ross, 2020).

With the Virtual Hearing Clinic (VHC) we provide a comprehensive, modular and distributed service platform aiming at utilizing ubiquitous mobile devices to foster targeted hearing health care. While existing approaches usually target one specific purpose or function (see above), the VHC offers a variety of potential functionalities for different users, such as patients, researchers, and audiologists, for use in individual hearing health care, research and clinical practice. The modular and web-based nature makes it easy to expand the framework by adding further modules and enables cross-platform compatibility. Functional modules have been developed for the different stages of hearing health care including self-administered diagnostics, treatment recommendation based on auditory profiling, benefit prediction, and provision of hearing support and self-adjustment for optimizing hearing support settings. VHC-based studies offer high flexibility in time and location, so that a larger number of potential participants can be addressed and barriers for participation are lowered. Consequently, VHC modules will be iteratively developed and advanced with a twofold vision: 1) The VHC as a mobile hearing health application and 2) as a research tool.

In this paper we introduce the concept and describe the architecture of the VHC. We give an overview of the current status of implementation regarding the underlying framework as well as functional modules. To exemplify the technical functionality, results from measurements with an available diagnostic module are shown and potentials and limitations are discussed.

## 2. Technical Description

### 2.1. General concept

The concept of the VHC is realized in a framework that integrates different functional modules in four areas: a) self-administered diagnostics and testing, b) auditory profiling, c) hearing support and benefit prediction and d) user-driven optimization of hearing support. These areas represent all potential stages of the hearing health patient journey. Due to the modularity the structure allows for continuous development and extension.

*a) self-administered diagnostics and testing:* Measurement and testing modules designed to be performed by the user without professional support and with basic consumer hardware, i.e., any mobile device with connected headphones. Measurements can include well established clinical procedures adapted for self-administered use as well as procedures specifically tailored for use in uncontrolled environments.

*b) auditory profiling:* Based on diagnostic data, users can be classified into distinct groups, i.e., profiles, characterizing hearing deficits beyond the audiogram. Data-driven auditory profiling uses machine-learning approaches to evaluate similarities and differences in diagnostic outcomes and relate those to potential action or treatment recommendations.

*c) hearing support and benefit prediction:* Illustration of the potential benefit of an intervention or the use of hearing support to be provided to the user. Integrating a software hearing aid can enable to experience the functions and practical values of technical hearing support. Another approach to illustrate the expected benefit of hearing aids is to provide a quantitative benefit estimation.

*d) user-driven optimization:* Methods and interfaces to enable user-driven adjustment and optimization of hearing support parameter settings without professional knowledge or support. The parameter spaces provided can be adapted based, e.g., on auditory profiles or computational modelling approaches.

### 2.2. Technical framework

The VHC is implemented as a browser-based application running on a server hosted by the Universität Oldenburg. The VHC can be accessed from any device with browser support and internet connection.

The design of the VHC framework follows a microservice architecture with a central orchestrator, the session application programming interface (API) (see Figure 1). The web application framework Flask [https://palletsprojects.com/projects/flask/] was used for the implementation of the main backend services and APIs: study setup (1), session handling and orchestration of the VHC hearing test modules (2), and saving study data (3). The functional hearing test modules (4) themselves are independent containerized web services that are connected to the framework via HTTP programming interfaces. All user connections are forwarded to the backend services via a reverse proxy (5) to bundle ports and to limit direct access to hidden setup service, data API, and session API routes. Each backend service contains their own domain specific databases.The study setup service offers a graphical webpage that enables researchers to set up an online hearing study from existing modules without the need of any programming. The study setup service also provides access control by creating unique identifiers for each test setup and access links for the participants. Each individual link will then load the configuration and redirect it to the session API. The session API offers an overview page, so participants can follow the progress along the whole test procedure. Experimental parameters such as calibration constants adjusting sound pressure levels across different modules are also administered by the session API. It also tracks the progress of each session, including starting a session, loading session and module configurations, navigating through the modules and caching the results of each module, and ensures that participants complete each module once as intended for the session. The data API manages the persistent storage of module-specific results and generic payloads. It serves as an additional layer between the modules and the actual SQL database preventing database queries in each module, which adds to the security of the framework.

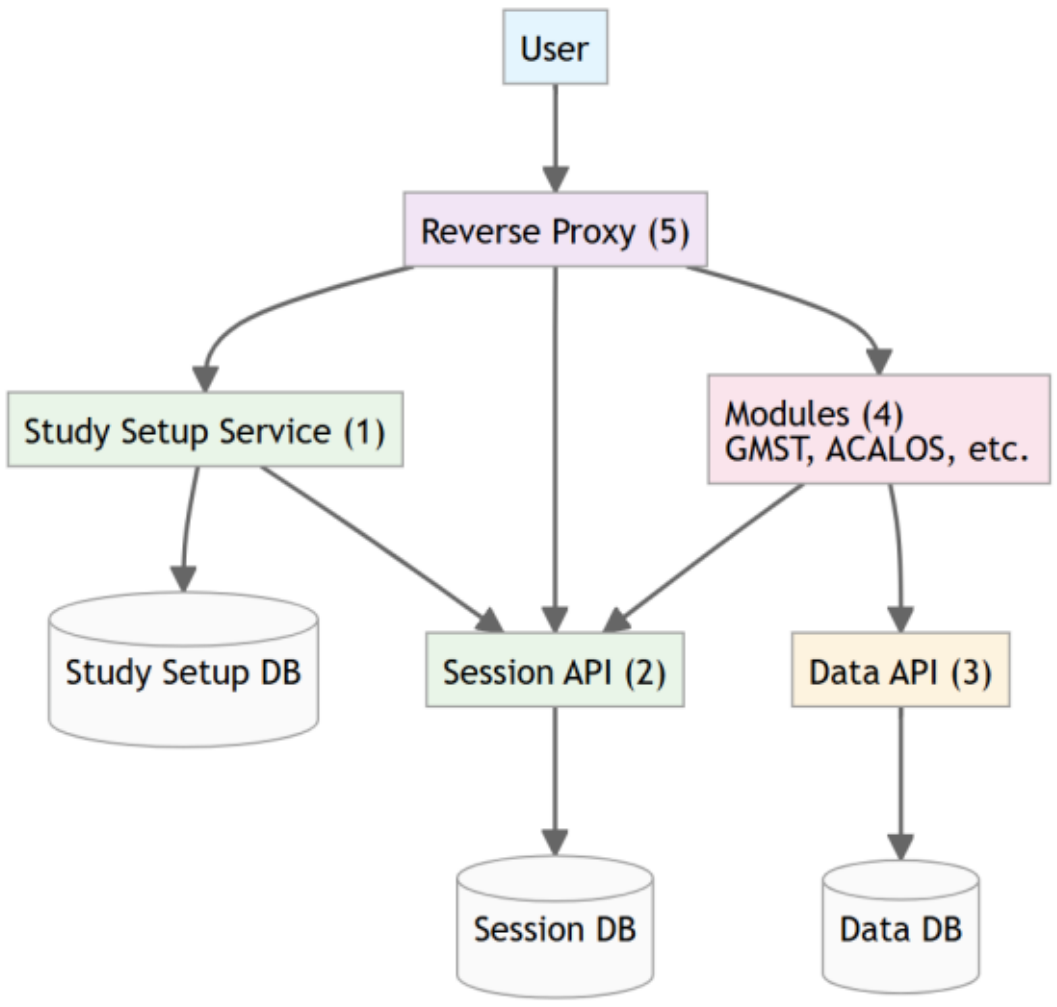

*Figure 1: Basic Structure of the VHC framework with three microservices for setup, session and data APIs and individual test modules. A reverse proxy bundles all microservice connections to one IP-address and port.*

### 2.1. Functional Modules

Functional modules are usually developed, implemented and validated in isolation before they are integrated into the VHC framework. So far, the main focus has been on diagnostics, which aims at collecting data from individuals, and is therefore essential for all further areas in the VHC. However, some work has also been done in the other areas. Table 1 gives an overview of functional modules that have been developed for the VHC. More information and further details can be found in the respective literature references.

| Measurement/Module | Functional description | Literature Reference |
|---|---|---|
| **Diagnostics** | | |
| German Matrix Sentence Test (GMST) | Determine speech reception threshold (SRT) in noise | (Saak, Kothe, et al., 2025; Schwarz et al., 2026) |
| Single Interval Up Down (SIUD) | Determine hearing thresholds (HT) in quiet | (Xu, Schell-Majoor, et al., 2024a) |
| Graded Response Bracketing (GRaBr) | Determine HT in quiet | (Xu, Hülsmeier, et al., 2024) |
| Adaptive categorical loudness scaling (ACALOS) | Determine loudness function | (Xu, Schell-Majoor, et al., 2024a) |
| ACALOS-based Calibration | Calibration estimation from ACALOS data | (Xu & Kollmeier, 2026) |
| Reinforced ACALOS | Determine loudness function and HT in quiet | (Xu, Schell-Majoor, et al., 2024b) |
| Tone in Noise | Determine tone-in-noise detection thresholds | (Isserstedt et al., 2026) |
| Specific Questionnaires | Collect additional information for specific purposes, e.g., identify people motivated to seek professional help ("help-seekers") | (Angonese et al., 2024) |
| **Profiling** | | |
| Auditory profiling | Deriving auditory profiles and classification of patients into profiles | (Saak et al., 2022) |
| Merging Profiles | Integrating further datasets to auditory profiles | (Saak, Oetting, et al., 2025) |
| **Benefit prediction** | | |
| Hearing aid benefit assessment | Assessment of perceived differences in distinct hearing aid settings | (Rullmann, 2025) |
| Cochlea Implant (CI)-Outcome Prediction | Predict postoperative speech recognition outcomes in CI recipients | (Demyanchuk et al., 2025) |
| **Optimization** | | |
| Self-Adjustment | Adjust individual hearing aid/support setting | (Gößwein et al., 2023) |

*Table 1: Overview over functional modules developed in the context of the VHC with literature references for more detailed description of the functionalities.*

## 3. Illustrative Example: Feasibility of a VHC-based audiogram

Audiometric thresholds, i.e., frequency-dependent absolute hearing thresholds, are the most used and established measurement in audiology. Therefore, in this chapter we illustrate the application of the VHC for obtaining an air conduction audiogram. With GRaBr (Xu, Hülsmeier, et al., 2024) an efficient and robust procedure has been developed and implemented as diagnostic module for the VHC. In this example, audiograms measured with the VHC are compared to clinical audiograms, i.e., the gold standard in audiology.

### 3.1. Participants

20 test subjects participated in the study which were recruited from the data base of the Hörzentrum Oldenburg gGmbH based on their audiogram (symmetric N2/N3 profile (Bisgaard et al., 2010)) and presumed ability to handle self-administered hearing tests on a tablet. There were 10 female and 10 male participants in an age range from 38 to 81 years with a mean age of 72.15 years. Participation was voluntary and all participants were paid on an hourly basis. The study was approved by the ethics committee of the University of Oldenburg (Drs. EK/2021/031-05).

### 3.2. Experimental setup

The reference setup comprised a calibrated clinical audiometer (Otometrics Madsen Astera) with audiometric headphones (Sennheiser HDA200) in an acoustically treated listening booth.
For the VHC setup the VHC was implemented on the Portable Hearing Laboratory (PHL), a hearing aid research platform consisting of mobile hardware (Pavlovic et al. 2018) featuring realistic ear-level devices and running the open Master Hearing Aid software (Kayser et al., 2022) for real-time, low-latency audio signal processing. As it is small, lightweight and flexible, field studies are a typical use case for the PHL. For this study, the PHL was connected with a tablet (Samsung Galaxy Tab A7 Lite) as the user interface and headphones (Audio-Technica ATH-M50x). The headphone output was equalized and calibrated using spatially diffuse speech-shaped noise. Further details can be found in (Rullmann, 2025).

### 3.3. Experimental procedures (GraBr)

The data was obtained during a lab appointment in the context of a larger study protocol focusing on comparing different hearing aid settings (Rullmann, 2025). First, hearing thresholds were measured with the reference setup (audiometer) at 0.125, 0.25, 0.5, 1, 2, 3, 4, 6 and 8 kHz. The measurement included also bone-conduction thresholds, which were not analyzed in the context of this study. If audiometric thresholds had been obtained for a participant within the last six months prior to the appointment, these data were taken as reference and this step was skipped. At the same appointment, but usually with other measurements in between, the audiogram was measured using the VHC setup with the GRaBr procedure at the frequencies 0.25, 0.5, 1, 2, 4 and 8 kHz in a quiet office room.

During GraBr measurements each test trial consists of the presentation of two tones at the current test frequency with different levels. The task of the participants is to indicate how many tones they heard (zero, one or two tones) on a graphical user interface displaying three respective buttons. Throughout the trials the levels of the tones are adjusted adaptively aiming at bracketing the hearing threshold between the tones.
In this study the probe tone was presented with a starting level of 65 dB and the initial level difference to the cue tone was +9 dB. With the adaptive procedure both tones are increased or decreased in level if the participant indicated zero or two heard tones, respectively. The starting step size for this level adjustment was 16 dB, which was decreased to 8 dB after the first and to 4 dB after the third reversal. The level difference between probe and cue tone was reduced to 6 dB when the participant first responded hearing a single tone and to 3 dB at the second occurrence of this response. Catch trials were implemented by including a 20 % chance for the tone with the higher level to be muted. If the response in these catch trials was 0 or 1 the level was increased or decreased, respectively, by half of the current step size. If the participant indicated 2 tones to be heard a false alarm was counted. The measurement was finished after a minimum of eight reversals and ten trials.

### 3.4. Results

Audiograms obtained with GRaBr implemented in the VHC were compared to those measured with an audiometer. As thresholds in GRaBr are output in dB SPL and the audiometer provides thresholds in dB HL, the hearing threshold levels from the audiometer ($L_{HT,HL}$) were converted to dB SPL according to Eq. 1 using the respective frequency specific reference equivalent threshold sound pressure level (RETSPL) values for the Sennheiser HDA200 from ISO 389-8 (International Organization for Standardization, 2004).

$$L_{HT,SPL} = L_{HT,HL} + L_{RETSPL} \qquad \text{(Eq. 1)}$$

The individual audiograms, i.e., hearing thresholds across certain frequencies, as measured with both methods are shown for the left and right ear of each participant in Figure 2. Visual inspection indicated generally good agreement. Especially the shapes of the individual audiograms were very similar with both methods.

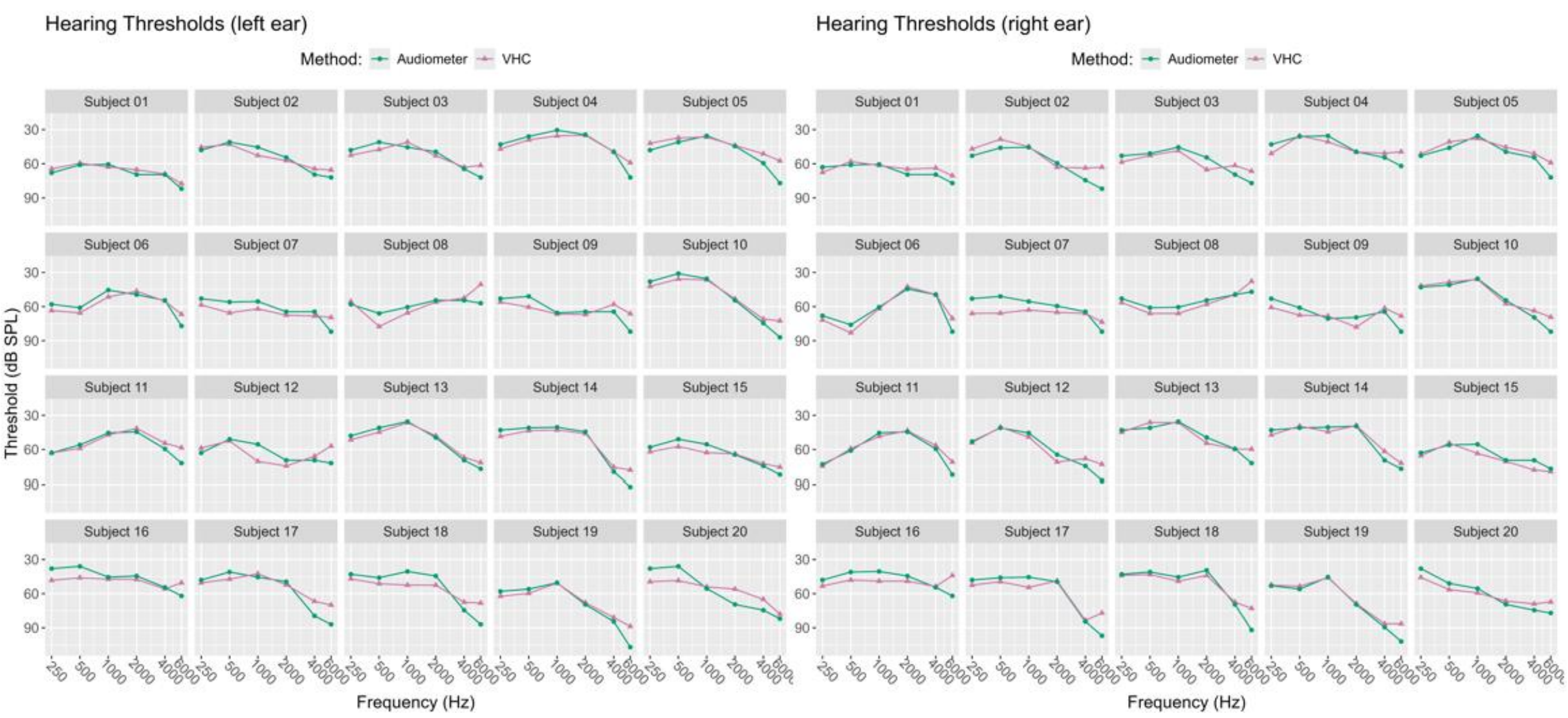

*Figure 2: Individual audiograms measured at six frequencies. Circles represent results from a standard audiometer in the lab (Audiometer)) and triangles results measured with the Virtual Hearing Clinic (VHC) for all participants. Note that the threshold values are given in dB SPL.*

The median threshold over all frequencies, subjects and ears was 55.5 dB SPL for the audiometer-based and 57.4 dB SPL for the VHC measurements. A Shapiro-Wilk test indicated that the differences between both methods over all frequencies and subjects significantly deviated from a normal distribution (Shapiro-Wilk test, $p < 0.05$). Hence, a two-sided paired Wilcoxon test was performed and did not find a significant difference between both methods (p-value = 0.58, V = 14562). Correlation analyses including all frequencies, participants and ears also showed high correlation between the methods with a Spearman correlation coefficient $r = 0.9$. Based on individual differences between both methods for all frequencies and ears, 60 % of the thresholds measured with the VHC were within 5 dB and 83 % within 10 dB of the thresholds measured with the clinical audiometer. Additionally, results were analyzed for the different frequencies, corresponding boxplots of frequency-dependent thresholds for both methods are shown in Figure 3. The Shapiro-Wilk test indicated that the assumption of normality was met for the differences between both methods for all frequencies ($p > 0.05$), thus paired t-tests where used. The corresponding p-values are shown in Table 2 together with intraclass (two-way model with absolute agreement for single measurements) and Pearson correlation coefficients as well as mean individual differences and root mean square errors (RMSE) for each frequency over all subjects and ears. Also, the percentages of VHC-based thresholds that were within 10 dB of thresholds obtained with the audiometer are given in the table.

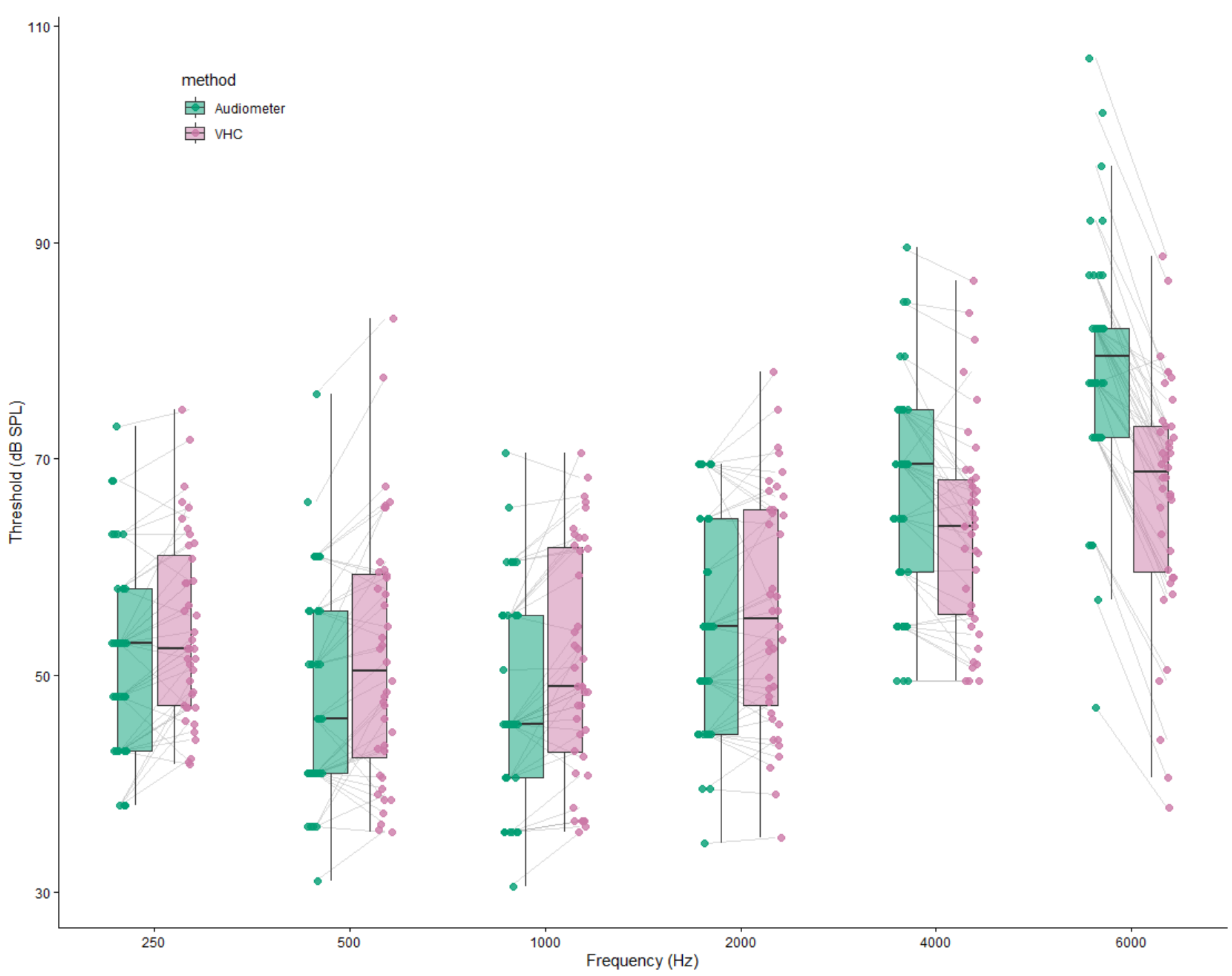

*Figure 3: Boxplots and individual data of hearing thresholds in dB SPL for different frequencies measured with a standard audiometer (Audiometer) and the Virtual Hearing Clinic (VHC). Data points from each subject are connected with grey lines. Boxes indicate the median and interquartile ranges over all subjects and ears.*

| | Mean difference (in dB) | RMSE (in dB) | ICC (two-way model) | R (Pearson) | p (paired t-test) | Within ±10 dB of reference |
|---|---|---|---|---|---|---|
| 250 Hz | 2.99 | 5.2 | 0.83 | 0.88 | < 0.05 | 92.5% |
| 500 Hz | 3.08 | 5.9 | 0.86 | 0.9 | < 0.05 | 92.5% |
| 1 kHz | 2.39 | 5.2 | 0.88 | 0.93 | < 0.05 | 95% |
| 2 kHz | 1.12 | 4.4 | 0.91 | 0.92 | 0.11 | 95% |
| 4 kHz | -3.28 | 5.2 | 0.87 | 0.92 | < 0.05 | 95% |
| 6 kHz | -12.17 | 13.1 | 0.57 | 0.9 | < 0.05 | 27.5% |

*Table 2: Mean individual differences between hearing thresholds measured with a clinical audiometer in the lab and measured with the VHC and root mean square errors (RMSE) for different frequencies. Additionally, correlation values (intraclass correlation coefficients (ICC), Pearson correlation coefficients (R)) relating thresholds from both methods are given as well as p-values testing the significance of the differences between both methods. Further, the percentages of thresholds measured with the VHC that are within 10 dB difference of those measured with the clinical audiometer (reference) are listed. Negative values indicate that reference thresholds are higher than VHC-based thresholds.*

While the t-tests indicated significant differences for all frequencies but 2 kHz, correlation values were high with ICC and R > 0.8 for all frequencies except 6 kHz. Mean differences and RMSEs on group level were around 3 dB and 5 dB, respectively, for all frequencies but 6 kHz. With more than 90%, the vast majority of measured thresholds for frequencies up to 4 kHz were within 10 dB of the reference thresholds. This indicates generally a good agreement between the methods for the frequencies from 250 Hz to 4 kHz with a slight trend towards increased thresholds values for the VHC, except for 4 kHz, where a trend towards lower VHC-based thresholds is observed. Looking at the results for 6 kHz, however, the agreement is much lower and VHC-based thresholds seem to be systematically decreased compared to the reference with a mean difference of -12.17 dB. RMSEs were considerably larger (13.1 dB) and the share of VHC-based thresholds lying within 10 dB of the audiometer-based reference thresholds much lower with only 27.5 %.

## 4. Discussion

In this paper we have given an overview of the general architecture of the VHC and presented several functional modules for the VHC that have been developed and tested with regards to their general feasibility. This includes self-administered diagnostic procedures that are available for data collection and self-testing. In this context we illustrated the feasibility of measuring audiometric thresholds with GRaBr, which generally showed good agreement with the clinical audiogram, especially the shape of the audiograms is very similar. This is also reflected in high correlation values (overall: spearman $r = 0.9$, frequency-specific: Pearson $r \geq 0.88$, ICC $\geq 0.83$, except for 6 kHz). Overall, 83 % of frequency-dependent thresholds measured with the VHC were within 10 dB of the clinical thresholds. Significant differences found with a t-test might partially be caused by the fact that the thresholds measured with the audiometer are only given in steps of 5 dB. One striking result is that differences in hearing thresholds at 6 kHz are much larger than at the other frequencies and VHC-based thresholds exhibit an offset towards lower thresholds compared to reference thresholds measured with an audiometer. Other studies have shown similar results when comparing self-administered audiometry to gold standard measurements: In (Adkins et al., 2024) averaged hearing thresholds (pure tone averages, PTA) obtained with a commercial app that was used with headphones delivered by the app provider reached correlations of up to $r = 0.8$ and 94% of PTAs were within 10 dB of those measured in a professional audiogram. When analyzing absolute threshold differences to the reference measurement, somewhat higher values of up to 96 % and 91% were found within 10 dB for app-based measurements with audiometric headphones in a quiet and noisy sound booth, respectively (Saliba et al., 2017). Another study comparing thresholds at different frequencies measured in a conventional or an automatic audiometry found correlations between $0.78 < r < 0.89$ for the different frequencies that overlap with the test frequencies in our study (Al-Maskari et al., 2026). In their study absolute median differences between both methods were mostly between 3.9 and 6.8 dB for test frequencies of 500 Hz to 4 kHz. By far the largest difference was found for the test frequency of 6 kHz with 13.9 dB which is in line with our findings. The decreased accuracy for higher frequencies is assumed to be caused by larger variances in headphone reproduction for frequencies of around 5 kHz and upwards, due to, e.g., interactions with the ear canal or placement issues. These variances are particularly relevant for non-professional headphones also for other VHC-measurements with narrowband high-frequency signals.

The use of mobile devices and usually non-professional headphones implies further limitations. One is the limited maximum output level, which might not be sufficient for people with more than a mild to moderate hearing loss. Therefore, VHC-based measurements with individual uncalibrated (consumer) hardware are not applicable for this user group, which is usually already aware of their impaired hearing. We recommend including questionnaires inquiring experienced and/or diagnosed hearing loss and use this as an inclusion criterion or to tag potentially invalid data. Another limitation concerns the increased uncertainty and variance caused by uncalibrated hardware. While a level adjustment should be performed before measurements with uncalibrated equipment, sound pressure levels will still vary between different hardware configurations. This is of minor importance in the case of measures based on level differences as it is the case in speech- or tone-in-noise measurements. However, to mitigate this problem one potential approach is to use more comprehensive measurements, such as adaptive categorical loudness scaling to estimate a device-specific calibration value (Xu & Kollmeier, 2026). Hence, the reliability of the calibration value is supposed to increase as the amount of audiologic measurements with appropriate plausibility checks are performed with the same uncalibrated device. For very short measurements with each participant and device, only calibration-independent audiological tests should therefore be employed (such as, e.g., speech-in-noise tests). A reliable absolute threshold estimate using uncalibrated hardware, however, appears only to be feasible in combination with other tests, such as, e.g., categorical loudness scaling.

## 5. Outlook

The VHC provides a modular platform for applications related to hearing health. While it has been shown to be feasible for data collection, basic diagnostics and adjustment of hearing support, the VHC will be continuously enhanced and modules will be refined aiming at a multifunctional tool for hearing research but also providing support throughout individual hearing health care trajectories.

Current work continuing the development of the VHC focusses on extending the validation database of the VHC diagnostics. This includes covering different degrees of hearing loss, combining different diagnostic modules and investigating the impact of specific influencing factors, e.g., different classes of hardware, calibration compensation procedures or decreased attention/motivation.

We also aim to utilize the VHC to collect larger and more diverse datasets for research and development as soon as possible. This can include remote web-based measurements with uncalibrated equipment as well as mobile measurements with calibrated equipment outside the lab. As audiological data is mainly collected for specific use cases and, especially in research, from rather small and biased samples, large-scale data collection utilizing the VHC can contribute to overcome this lack of "big data" in audiology. These data can further be combined with data from other sources and provide the basis for developing data-driven applications improving individual hearing health care. This also refers to refining and evaluating functional modules of the VHC that are based on machine learning, e.g., auditory profiling.

**Acknowledgments**

This work was funded by the Deutsche Forschungsgemeinschaft (DFG, German Research Foundation) under Germany's Excellence Strategy – EXC 2177/1 – Project ID 390895286.

The authors would like to thank Chen Xu for providing the code for the GRaBr measurements as well as Chiara Haf and Kerstin Sommer for their help in data collection.

**Disclosure statement**

The authors report there are no competing interests to declare.

**Declaration of generative AI use**

The authors report that Chat GPT (GPT 5.3 and 5.5) was used for coding assistance to conduct statistical analyzes and plotting of the data in R (https://www.R-project.org/).